# NAAS: Negotiation Automation Architecture with Buyer's Behavior Pattern Prediction Component

Debajyoti Mukhopadhyay[1], Sheetal Vij[2], Suyog Tasare[2]

Department of Information Technnology[1]
Department of Computer Engineering[2]
Maharashtra Institute of Technology
Pune 411038, India
{debajyoti.mukhopadhyay, sheetal.sh, suyog59}@gmail.com

**Abstract:** In this era of "Services" everywhere, with the explosive growth of E-Commerce and B2B transactions, there is a pressing need for the development of intelligent negotiation systems which consists of feasible architecture, a reliable framework and flexible multi agent based protocols developed in specialized negotiation languages with complete semantics and support for message passing between the buyers and sellers. This is possible using web services on the internet. The key issue is negotiation and its automation. In this paper we review the classical negotiation methods and some of the existing architectures and frameworks. We are proposing here a new combinatory framework and architecture, NAAS. The key feature in this framework is a component for prediction or probabilistic behavior pattern recognition of a buyer, along with the other classical approaches of negotiation frameworks and architectures. Negotiation is practically very complex activity to automate without human intervention so in the future we also intend to develop a new protocol which will facilitate automation of all the types of negotiation strategies like bargaining, bidding, auctions, under our NAAS framework.

## 1. Introduction to Negotiation

Negotiation is the process between two or multiple entities where everybody comes to some useful consensus or agreement as a result. Like it or not everybody is a negotiator in some ways without even knowing it. We did it as kids for trading toys, cards and still we do it for the raise in salary, purchasing things in our personal lives. Many people try to avoid this blatant negotiation procedure consciously because they don't like it but end up in either negotiating or losing in the bargain. Now the question is how to automate this process using the latest technologies and advancements in computing. Here are some key concepts behind the theory of negotiation [6, 7, and 8].

### 1.1 Negotiation Types

1. Distributive or fixed pie negotiation. It involves people who have never had a previous interactive relationship neither they are likely to do so in near future. Everybody gets a fixed pie. 2. Integrative negotiation or everybody wins something or win-win scenario. This means to join several parts into whole. This needs some cooperation and higher degree of trust from every entity in negotiation. Ideally it is difficult to achieve and most difficult to automate because trust and forming relationships on that is human characteristic and difficult to implement in machines and computing [9].

### 1.2 Negotiation Tactics

The level of detail in the best negotiator actually understands the human mind and how to use this in different voice tones and expressions for the best possible negotiation outcome for all the parties, ideally. A few common tactics that are used in negotiation are outright refusal, conditioning, calling bluffs [10].

### 1.3 Negotiation protocol types

It is the set of rules that govern the interaction between participants. Depending upon the types the negotiation can be categorized as bidding, auction, bargaining. [11]

In this paper we are proposing a new architecture and a framework for automating the negotiation process. The key feature in this framework is a component for prediction or probabilistic behavior pattern recognition of a buyer, along with the other classical approaches of negotiation frameworks and architectures.

## 2. Related Work

Substantial research work has been done on various negotiation protocols, languages and set of parameters, frameworks on each of the negotiation types like auctioning, bidding, bargaining, simple request and response methods etc. We reviewed the following frameworks and architectures to come up with certain conclusions and assessed if any provision is given in those for observing the buyer's behavior.

Hudert, Guido and Ludwig [1] have defined a framework for augmenting WS-Agreement by Open grid Forum (OGF) standard which actually defines an XML based structural definition of Service Level Agreements [2]. This framework is proposed only in relation with WS-Agreement protocol where parties interested in negotiating an agreement first run the negotiation Meta protocol to establish which negotiation protocol is used. [1] Subsequently, the protocol is executed to determine the resulting, negotiated WS-Agreement document. Finally, winner is determined and acceptance, rejection is performed again according to the WS-Agreement protocol standard (Fig. 1).

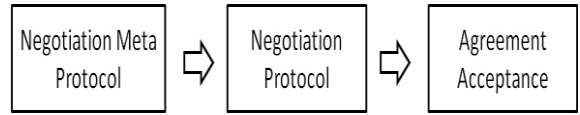

**Fig. 1.** Agreement creation process

Tung Bui and Gachet [3] found out that web services (using UDDI, WSDL, SOAP, and XML) can be used as a market broker, to help in discovering the supply/demand, arbitrate the pricing, find the most appropriate service for a given request, to modify the request and services and generate the contract. They have given the basic architecture as shown in Fig. 2.

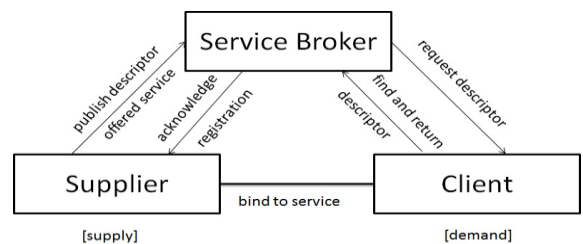

**Fig. 2.** Electronic market for web services

In this framework the authors have used BPEL4WS[4,5] and WSCI and proposed following support roles as information, Communication, Negotiation, Bargaining, Execution of transactions , all of them with steps in pre-transaction, transaction and post transaction phases in electronic market. Authors have given detailed diagrams for the seven web services like topology of web services for negotiation and bargaining, service discovery, Adaptation and pricing, Service ranking, service bargaining, best price adaptation, contract composition, here the client is seen as a negotiation manager. This architecture has its limitations like assumption of trust and there is no competitive strategy to distort the cooperative spirit of e-market. There is no fixed pricing, brokerage and addressing of services, no technical requirements as security and logging mechanisms discussed.

Bin Wu and Chaozhen Guo [12] present a new web service negotiation mechanism and new web service composition coordinated negotiation architecture to solve the problems which mainly occur in such architectures of web service composition based on agent. The problems like unreasonable use of time and data link resources in the condition of multi-negotiation concurrency which leads to inefficiency of negotiation, lack of effective processing when confronting negotiation failure. Here the authors are applying asynchronous communication theory to the process of Web Service

Negotiation and extend effective processing in case of negotiation failure. There are disadvantages as overhead of time spent in processing failure may be more than the benefit it brings, if the bad status of network (Fig. 3 and Fig. 4).

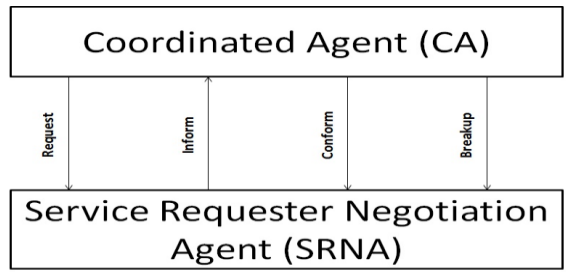

**Fig. 3.**   Original web service composition coordinated negotiation architecture

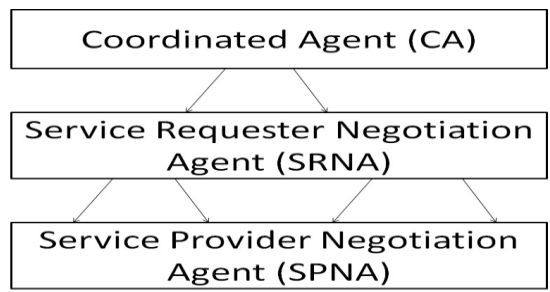

**Fig. 4.**   Improved web service composition coordinated negotiation architecture (IWSCCNA)

It divides NA into SPNA and SRNA. Here a pseudo algorithm of process handling negotiation failure is given. So the research focuses only on elaborating architecture (Improved Vs OWSCCNA) for proper processing in case of negotiation failure in web services.

Jin and Segev [13] have proposed a framework for negotiation processes that provide a consistent model for supporting a comprehensive range of negotiations in dynamic next generation e-business environment. It has five components negotiation requirements, negotiation structure, negotiation process, negotiation protocol and strategy. The authors have described each of their framework components in stepwise details and claim that this model is the most flexible and practical where protocol and strategy are separated in the design, as shown in Fig. 5

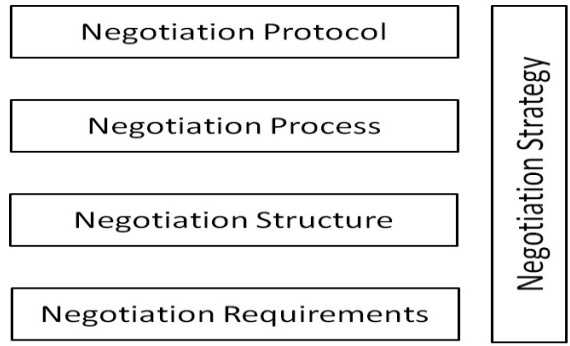

**Fig. 5.** Negotiation Framework

Chhetri, Lin, Goh, jun, Jian, Kowalczyk [14] have proposed an agent based coordinated negotiation architecture to ensure collective functionality, end to end QoS and coordination of complex services and they describe how this architecture can be used in different application domains and also how the negotiation system on the service provider's side can be implemented both as an agent based negotiation as well as a web service based negotiation system. This work is based on ASAPM, adaptive agreement and process management, aims at developing intelligent agent based techniques and tools to facilitate the adaptive service management and process management. The overall architecture of ASAPM includes following four components (Fig. 6).

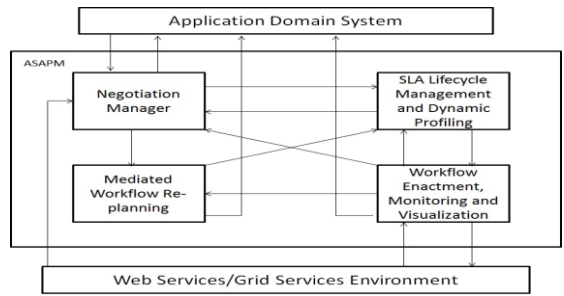

**Fig. 6.** ASAPM

Coordinated negotiation architecture is proposed on the above as shown in Fig. 7.

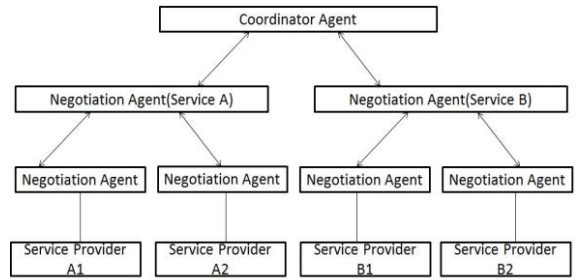

**Fig. 7.** Coordinated Negotiation Architecture

This two level approach enables the reuse of this architecture in any application domain. This paper does not contain the decision making strategies both at the negotiation level and at the coordination level.

All these architectures are mostly implemented in FIPA compliant JADE Agent Framework and WS2JADE toolkit which enables the integration of JADE agents and web services.

Stanley, Huang, Yihua, Haifei Li, Wang, Liu, Lee, Lam [15] have presented the design and implementation of a replicable, internet based negotiation server for conducting bargaining type negotiations between enterprises involved in e-commerce and e-business where enterprises can be buyers and sellers of product/services or participants of a complex supply chain involved in purchasing, planning and scheduling. The use of negotiation servers to conduct automated negotiation has been demonstrated by the authors. A content specification language for information registration, a negotiation protocol and its primitive operations, an automated negotiation process, a cost benefit decision model and the architecture of an implemented system have been described in this. This is based on object oriented, active database technology in contrast to the existing systems which are based mostly on distributed agent technology. Their negotiation server is analogous to web server which provide following negotiation services as a registration service, a negotiation proposal processing service, an event and rule management service, a cost-benefit analysis service. The system architecture consists of following components, external to WAN components as shown in Fig. 8.

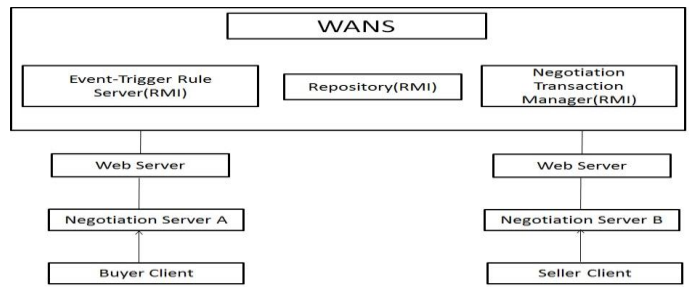

**Fig. 8.** System Architecture

Cao, Chi, Liu [17] have also described an automated negotiation architecture based on software multi agents using SOA (Service oriented architecture) and web services technology. Here is a reference to Negotiation as a Service where the authors observed that in businesses, negotiation process is better given as a service, not as software and more suitably the entities can be seen in a role of a service provider. Authors correlate it with SaaS (software as a service) where software providers deploy application software on their own servers and customers search, access software services via Internet, consume services as per demand and pay the software providers based on time and number of services consumed. So the users buy software service as an alternative to permanent license and SaaS subverts the traditional life cycle of software i.e. design- development-testing-installation. User need not concern about the upgrade and maintenance of software. Here the authors have given a six layered architecture for the automated negotiation as in Fig. 9.

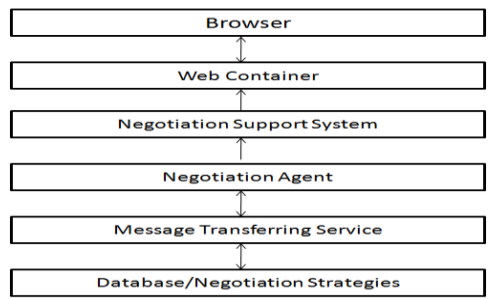

**Fig. 9.** Architecture for Automated Negotiation

Though this prototype has been developed using SOA and UDDI, WSDL, XML technologies , it could not be a standard for practical implementation of negotiation automation due to many pervasive issues and conflicts in standards and overhead due to parsing of SOAP messages and XML documents on every exchange. Precisely in all the above architectures, even if they can be converted into a prototype of automated negotiation, they cannot be a fixed standard for the practical implementation of the same. Also there is no consideration of understanding and predicting the buyer's behavior in all the above architectures, which is if added it can be a very sophisticated and futuristic approach of negotiation automation.

## 3. NAAS Automation Architecture

We are proposing an automated negotiation system in the form of web service calling it as NAAS (Negotiation as a Service). As we saw in survey, using SOA (Service oriented architecture) we can make the system running in any place as a service node that is integrated with third party e-commerce platforms so the system can play role of negotiation service provider in the real business environment. The benefits are , we can obtain stable visiting quantity, maintenance and upgrade of system can be completed on the server independently, saving human and material resources, automated negotiation system can make use of the existing basic facilities provided by e-commerce platform i.e. security, authentication, transaction management etc. ,saving costs of development.

With NAAS we are trying to overcome the issues we saw in the previous architectural styles. Specifically we are trying to add a module which can be developed on a strong algorithmic base i.e. using Association rules or Markov models to predict and then further analyzing the buyer's behavior for making the process of negotiation automation complete. The main reason for buyer's behavior prediction module in the basic design is to make the negotiation system intelligent, sophisticated and futuristic so that a better standard for negotiation automated can be created and existing designs can be enhanced.

NAAS Architecture Components: 1.Service registry (databases types, directory) 2.Negotiation support system (Negotiation support system provider, Negotiation service requestor) 3.Protocol on internet module, MTS and service discovery by using web services on UDDI, HTTP, SOAP 4.Advertisement publishing from provider/seller 5.Negotiation service requestor/ buyer 6.Buyer's behavior pattern prediction (proposed in auction, in bargaining, in bidding for further research by the authors) 7.Business logic module and agent management module (external to NAAS) 8.Strategies, decision modules (external to NAAS).

Working: The seller will publish its information about product and the prices etc. On the service registry via web services and web container to which a negotiation support system will interact on some negotiation protocol or all the existing protocols. A MTS is on the internet to transfer the requests from buyers as well as product information from the sellers. A database can be maintained for all these service related queries and information with different ontology, if this architecture is considered in details. In our buyer's behavior prediction component, some key features like age, gender, culture, type of product, feedback can be taken as input to the system for the decision making and predicting the buyer's choices, behavior according to the region, country and states. This will help us in tapping the particular market and applying further strategies in negotiation system according to the buyer's behavioral aspects for all the types in negotiation like bidding, bargaining and auctions for the desired product on sale (Fig. 10).

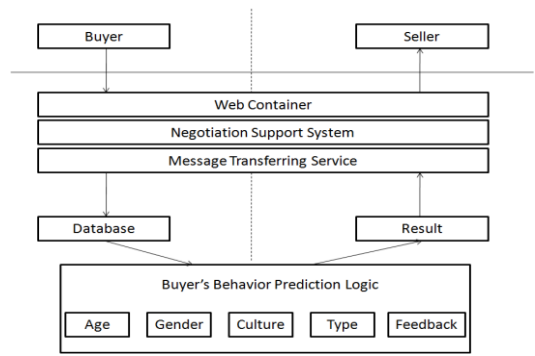

**Fig. 10.** NAAS Architecture

## 4. Discussion on buyer's behavior prediction component

Due to addition of the buyer's behavior pattern prediction, this architecture can be a flexible, reliable because if in advance of any negotiation process in a multiple buyers and sellers scenario on web based automated negotiation system, we are able to analyze how many potential buyers can be in this negotiation process and what kind of negotiation behavior the particular set of buyers go into. We are proposing a buyer's behavior prediction component as shown in Fig. 11.

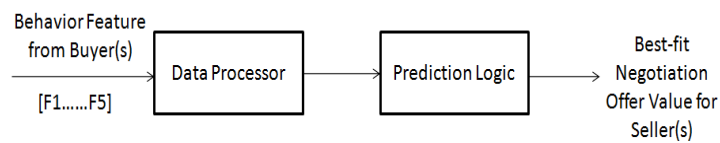

**Fig. 11.** Buyer's behavior prediction component

First, data processor processes the input and then forwards the required values to the prediction logic. Our proposed prediction logic computes the offer values according to the feature which is submitted by the buyer.

$$\text{Offer Value (OV)} = \frac{\text{Feature [F1.....F5]}}{\text{Count of Buyer(s)}} * 100$$

According to the present proposed prediction logic, we can generate such five types of offer values according to the particular feature (e.g., age, gender, culture, type of product, buyer's demand or feedback)

For e.g. Feature (F1) = Age then we can divide this into 3 categories like Age[10 - 30] , [30 - 50 ], [50 – 70], we can allocate a weighted value for each of these age groups which will be divided by count (C) of buyer(s) to calculate the percentage. This percentage can be used by seller(s) to negotiate on the Offer Value (OV) of that particular feature (F1...F5). Please see Fig. 12.

| Feature No. | Feature | Range |
|---|---|---|
| 1 | Age | [10-30], [30-50], [50-70] |
| 2 | Gender | Male, Female |
| 3 | Culture | Region/Country |
| 4 | Type of Product | Product Categories |
| 5 | Feedback | Buyer's Feedback |

**Fig. 12.** Feature Categories

This leads to a very successful implementation of various negotiation strategies (auction, bidding, bargaining) etc. There is no existing architecture on negotiation which facilitates the intelligent buyer's behavior prediction in automated negotiation. We are working on the survey of appropriate existing algorithms for the prediction.

## 5. Conclusion and Future Work

The overall survey of existing architectures leads us to come up with shortfalls and requirements in negotiation automation. NAAS can provide the advantages and further applications related to this. The authors of this manuscript are trying to assess the work done in this area and come up with some conclusions about how to construct as well as deploy the buyer's behavior prediction component for the e-negotiation system. We intend to find out an appropriate algorithmic base to implement this scheme. We also intend to create a new negotiation protocol for our NAAS framework in the next phase of work.